\documentstyle[12pt,epsf]{article} 

\def\lsim{\mathrel{\lower2.5pt\vbox{\lineskip=0pt\baselineskip=0pt 
           \hbox{$<$}\hbox{$\sim$}}}} 
\def\gsim{\mathrel{\lower2.5pt\vbox{\lineskip=0pt\baselineskip=0pt 
           \hbox{$>$}\hbox{$\sim$}}}} 

\def\da{\dot{a}}
\def\dda{\ddot{a}}
\def\db{\dot{b}}
\def\ddb{\ddot{b}}
\def\dx{\dot{X}}
\def\ddx{\ddot{X}}
\def\dy{\dot{Y}}
\def\ddy{\ddot{Y}}
\def\dz{\dot{Z}}
\def\ddz{\ddot{Z}}

\def\g{\gamma}
\begin{document} 
\begin{flushright}
DPNU-03-13\\ hep-th/0305130
\end{flushright}

\vspace{10mm}

\begin{center}
{\Large \bf 
 On the solutions to accelerating cosmologies}

\vspace{20mm}
 Masato Ito 
 \footnote{E-mail address: mito@eken.phys.nagoya-u.ac.jp}
\end{center}

\begin{center}
{
\it 
{}Department of Physics, Nagoya University, Nagoya, 
JAPAN 464-8602 
}
\end{center}

\vspace{25mm}

\begin{abstract}
 \baselineskip=6mm
 Motivated by recent accelerating cosmological model,
 we derive the solutions to vacuum Einstein equation in
 $(d+1)$-dimensional Minkowski space with $n$-dimensional
 hyperbolic manifold.
 The conditions of accelerating expansion are given in such a set up.
\end{abstract} 

\newpage 
 \baselineskip=7mm

 Recently, Townsend and Wohlharth discovered a solution of the
 four-dimensional accelerating cosmologies with compact hyperbolic
 manifold \cite{Townsend:2003fx}.
 By avoiding the assumptions of a no-go theorem
 \cite{Maldacena:2000mw}, it was pointed out that
 the accelerating phase arises in a interval.
 The features of the model are that the four-dimensional metric is
 Einstein frame and the compact hyperbolic manifold is time-dependent.
 Before \cite{Townsend:2003fx},
 the investigations of FRW universes with closed hyperbolic internal space
 have been widely performed
 \cite{Starkman:2000dy,Kaloper:2000jb,Starkman:2001xu,
 Silva:2002nq,Nasri:2002rx}.
 Furthermore it has been shown that the model is closely connection with
 the S-branes which correspond to time-dependent solutions of
 supergravity with $p$-form 
 \cite{Gutperle:2002ai,Kruczenski:2002ap,Ohta:2003uw,
 Ohta:2003pu,Ohta:2003ie,Emparan:2003gg,Wohlfarth:2003ni}.
 Thus it is expected that the accelerating universe supported by
 astronomical observations may be explained by string/M-theory.

 From the theoretical point of view, we are interested in the higher
 dimensional cosmological model motivated by the model of
 Ref.\cite{Townsend:2003fx}.
 In this letter, we derive the solutions to the vacuum Einstein equation
 in time-dependent ${\cal M}_{d+1}\times H_{n}$, where ${\cal M}_{d+1}$
 and $H_{n}$ are $(d+1)$-dimensional Minkowski and $n$-dimensional
 hyperbolic manifold, respectively.
 Furthermore we investigate the conditions of accelerated expansion
 of the present model.
 
 Adopting the Einstein frame metric for $d\geq 3$, we can take ansatz for
 time-dependent metric
 \begin{eqnarray}
 ds^{2}=b^{\frac{4}{d-1}}(t)ds^{2}_{d+1}
 +b^{-\frac{4}{n}}(t)d{\hat{s}}^{2}_{n}
 \label{eqn1}\,,
 \end{eqnarray}
 where
 \begin{eqnarray}
 ds^{2}_{d+1}=-a^{2d}(t)dt^{2}+a^{2}(t)dx^{2}_{d}\,,\label{eqn2}
 \end{eqnarray}
 where $a(t)$ corresponds to the scale factor of $d$-dimensions.
 For the $n$-dimensional hyperbolic manifold, the Ricci tensor is
 given by
 \begin{eqnarray}
 \hat{\cal R}_{ab}=-(n-1)g_{ab}\,.\label{eqn3}
 \end{eqnarray}
 The Ricci tensors for metric (\ref{eqn1}) are evaluated as follows,
 \begin{eqnarray}
 {\cal R}_{tt}&=&-\frac{2}{d-1}\frac{\ddb}{b}
          -2\left(\frac{1}{d-1}+\frac{2}{n}\right)\frac{\db^{2}}{b^{2}}
          -d\frac{\dda}{a}+d^{2}\frac{\da^{2}}{a^{2}}
          \nonumber\\
 {\cal R}_{\mu\nu}&=&a^{2(1-d)}\left\{
          \frac{2}{d-1}\left(\frac{\db}{b}\right)^{\dot{}}
          +\left(\frac{\da}{a}\right)^{\dot{}}
          \right\}g_{\mu\nu}
          \nonumber\\
 {\cal R}_{ab}&=&
 \left[
 -\frac{2}{n}b^{-4\left(\frac{1}{n}+\frac{1}{d-1}\right)}
  a^{-2d}\left(\frac{\db}{b}\right)^{\dot{}}
 -(n-1)\right]g_{ab}
 \label{eqn4}
 \end{eqnarray}
 where $\mu,\nu$ and $a,b$ represent ${\cal M}_{d+1}$ and $H_{n}$,
 respectively.
 The dot denotes the derivative with respect to the time $t$.
 In order to simplify the vacuum Einstein equation for metric ansatz,
 we define variables
 $X$ and $Y$ by
 \begin{eqnarray}
  b&=&e^{X}\nonumber\\
  a&=&e^{Y}\,.\label{eqn5}
 \end{eqnarray}
 Using the new variables, the equations of motion can be rewritten as
 \begin{eqnarray}
 -\frac{2}{d-1}\ddx-4\left(\frac{1}{d-1}+\frac{2}{n}\right)\dx^{2}
 -d\ddy+d(d-1)\dy^{2}&=&0
 \,,\label{eqn6}\\
 \frac{2}{d-1}\ddx+\ddy&=&0\,,\label{eqn7}\\
 \ddx+\frac{1}{2}n(n-1)
 e^{4\left(\frac{1}{d-1}+\frac{1}{n}\right)X+2dY}
 &=&0\,.\label{eqn8}
 \end{eqnarray}
 Performing the integral of (\ref{eqn7}), and we have
 \begin{eqnarray}
 \frac{2}{d-1}X+Y=ft+g\label{eqn9}\,,
 \end{eqnarray}
 where $f$ and $g$ are integration constants.
 Substituting (\ref{eqn9}) into (\ref{eqn6}) and (\ref{eqn8}), the
 equations of motion can be written in terms of $X$
 \begin{eqnarray}
 \ddx+2\frac{n-1}{n}\dx{}^{2}-2fd\dx
 +\frac{1}{2}d(d-1)f^{2}=0\label{eqn10}\\
 \ddx=-\frac{1}{2}n(n-1)e^{2gd}e^{-4\frac{n-1}{n}X+2dft}\label{eqn11}\,,
 \end{eqnarray}
 In order to solve (\ref{eqn11}), we define $Z$ by
 \begin{eqnarray}
 Z=-4\frac{n-1}{n}X+2dft\,.\label{eqn12}
 \end{eqnarray}
 Consequently, the second order differential equation can be obtained 
 \begin{eqnarray}
 \ddz=2(n-1)^{2}e^{2gd}e^{Z}\,.\label{eqn13}
 \end{eqnarray}
 Performing the integral of the above equation, we have
 \begin{eqnarray}
 \dz{}^{2}=4(n-1)^{2}e^{2gd}e^{Z}+h\,,\label{eqn14}
 \end{eqnarray}
 where $h$ is positive integration constant.
 Therefore, the solution to the $Z$ is given by
 \begin{eqnarray}
 e^{Z}=\frac{h}{4(n-1)^{2}}\frac{e^{-2gd}}
 {\sinh^{2}\frac{\sqrt{h}}{2}\left(t+t_{0}\right)}\,,
 \label{eqn15}
 \end{eqnarray}
 where $t_{0}$ is the integration constant.
 Furthermore, (\ref{eqn12}) can be rewritten in terms of $Z$
 \begin{eqnarray}
 2n\ddz-n\dz{}^{2}+4f^{2}d\left(n+d-1\right)=0\,.
 \label{eqn16}
 \end{eqnarray}
 Substituting (\ref{eqn13}) and (\ref{eqn14}) into (\ref{eqn16}),
 the relation between the integration constants $h$ and $f$ can be
 expressed as
 \begin{eqnarray}
 h=4\frac{d(n+d-1)}{n}f^{2}\,.\label{eqn17}
 \end{eqnarray}
 Here we simply take $f=1$, $g=0$, $t_{0}=0$.  
 Combining with (\ref{eqn9}), (\ref{eqn12}) and (\ref{eqn15}),
 and we have
 \begin{eqnarray}
 b&=&e^{X}=e^{\frac{nd}{2(n-1)}t}
           \left(\frac{\g}{(n-1)\sinh \g|t|}\right)^{-\frac{n}{2(n-1)}}
    \label{eqn18}\\
 a&=&e^{Y}=e^{-\frac{n+d-1}{(n-1)(d-1)}t}
           \left(\frac{\g}{(n-1)\sinh \g|t|}\right)
    ^{\frac{n}{(d-1)(n-1)}}
    \label{eqn19}
 \end{eqnarray}
 where
 \begin{eqnarray}
 \g=\sqrt{\frac{d(n+d-1)}{n}}\,.\label{eqn20}
 \end{eqnarray}
 Note that the argument of the hyperbolic sine can be taken to be
 positive in order that the solutions are real.

 For $d=3$, obviously, (\ref{eqn18}) and (\ref{eqn19}) are completely
 consistent with the metric in Ref.\cite{Townsend:2003fx}.
 Furthermore it is considered that the results obtained here are
 included in S-brane solutions in arbitrary dimensions derived by
 N. Ohta \cite{Ohta:2003uw}.

 Below we shall investigate expansion and acceleration in the case of
 $d\geq3$.
 From the metric (\ref{eqn2}), the proper time $\eta$ for
 $(d+1)$-dimensional world is given by $d\eta=a^{d}dt$.
 The condition of expanding $(d+1)$-dimensional universe, $d a/d\eta >0$,
 leads to 
 \begin{eqnarray}
 M(t)=1+\sqrt{\frac{nd}{n+d-1}}\coth
        \left(\sqrt{\frac{d(n+d-1)}{n}}\;t\right)<0\,.\label{eqn21}
 \end{eqnarray}
 The above inequality is satisfied for negative $t$ since the
 coefficient of hyperbolic cotangent is always larger than one.

 Moreover the condition of accelerating $(d+1)$-dimensional universe,
 $d^{2} a/d\eta^{2} >0$, leads to
 \begin{eqnarray}
 M^{2}(t)<
 \frac{d(n-1)}{n+d-1}
 \frac{1}{\sinh^{2}\left(\sqrt{\frac{d(n+d-1)}{n}}\;t
 \right)}\,.\label{eqn22}
 \end{eqnarray} 
 The accelerating phase arises if the conditions (\ref{eqn21}) and
 (\ref{eqn22}) are simultaneously satisfied at negative $t$.
 As for the asymptotic relationship between time $t$ and 
 proper time $\eta$,
 note that $t\rightarrow -\infty$ corresponds to $\eta\rightarrow +0$ while
 $t\rightarrow -0$ corresponds to $\tau\rightarrow \infty$.
 Furthermore the asymptotic behavior of scale factor for $\eta$
 is expressed as
 \begin{eqnarray}
 a(\eta)\sim\left\{
 \begin{array}{ll}
 \eta^{\frac{1}{d}}
 & {\rm for\;}\eta\rightarrow +0\;
 (t\rightarrow -\infty) \\ \\
 \eta^{\frac{n}{n+d-1}}
 & {\rm for\;} \eta\rightarrow \infty\;
 (t\rightarrow -0)
 \end{array} 
 \right.\label{eqn23}\,.
 \end{eqnarray}
 For small $\eta$, note that scale factor is independent of $n$.
 
 The interval of accelerating expansion is determined by
 $d^{2}a/d\eta^{2}=0$.
 Therefore we get
 \begin{eqnarray}
 t_{1}&=&-\sqrt{\frac{n}{d(n+d-1)}}\log
 \frac{\sqrt{n-1}(\sqrt{d}+1)}{\sqrt{nd}-\sqrt{n+d-1}}\nonumber\\
 t_{2}&=&-\sqrt{\frac{n}{d(n+d-1)}}\log
 \frac{\sqrt{n-1}(\sqrt{d}-1)}{\sqrt{nd}-\sqrt{n+d-1}}
 \label{eqn24}\,,
 \end{eqnarray}
 where $t_{1}$ and $t_{2}$ are start and end of acceleration,
 respectively.
 Using the above results, we can estimate the e-foldings number in
 an interval of acceleration, namely,
 $\log(a(t_{2})/a(t_{1}))={\cal O}(1)$.
 As indicated in Ref.\cite{Chen:2003ij}, since the e-foldings number is of
 order one, it is questionable whether the present model is worthy of
 inflation model.
 It is expected that models or mechanisms for enlarging 
 e-foldings number are proposed in the future. 

 In summary, we derived the solutions to vacuum Einstein equation
 in ${\cal M}_{d+1}\times H_{n}$ with
 Einstein frame metric for $d\geq3$, triggered by recent accelerating
 cosmological model of Ref.\cite{Townsend:2003fx}.
 We provided the conditions of accelerating expansion and
 the asymptotic behaviors of scale factor.
 Moreover, the start and end of acceleration are given and it was pointed
 out that the e-foldings number during the interval is approximately of
 order one for proper values of $n,d$. 

{\bf Acknowledgements:}

 We thank N. Ohta for pointing out errors in the earlier version
 of the paper and useful correspondence.

{\bf Note added:}
 After we submitted this paper to the hep archives, we learned the higher
 dimensional accelerating cosmologies with product spaces proposed by
 C-M. Chen et al.\cite{Chen:2003ij}. 
 It was pointed out that the model considered here is related to the
 S-branes by N. Ohta \cite{Ohta:2003uw}.  
%

\end{document}